\PassOptionsToPackage{table}{xcolor}
\documentclass[sigconf, nonacm]{acmart}

\usepackage{balance}
\usepackage{subcaption}
\usepackage{enumitem}

\usepackage{amsmath,amsfonts}
\usepackage{graphicx}
\usepackage{textcomp}
\usepackage{siunitx}

\usepackage{url}
\usepackage{multirow}
\usepackage{xspace}
\usepackage[ruled,linesnumbered,vlined]{algorithm2e}
\usepackage{amsmath}

\SetKwInput{KwInput}{Input}

\newcommand{\algo}{ELSAR\xspace}
\newcommand{\R}{\textsuperscript{\textregistered} }
\newcommand{\TM}{\textsuperscript{\texttrademark} }

\begin{document}
\title{Parallel External Sorting of ASCII Records Using Learned Models}

\author{Ani Kristo}
\orcid{0000-0002-1825-0097}
\affiliation{%
  \institution{Brown University}
}
\email{ani@brown.edu}

\author{Tim Kraska}
\affiliation{%
  \institution{MIT}
}
\email{kraska@mit.edu}

\begin{abstract}
External sorting is at the core of many operations in large-scale database systems, such as ordering and aggregation queries for large result sets, building indexes, sort-merge joins, duplicate removal, sharding, and record clustering.
Unlike in-memory sorting, these algorithms need to work together with the OS and the filesystem to efficiently utilize system resources and minimize disk I/O.

In this paper we describe \algo: a parallel external sorting algorithm that uses an innovative paradigm based on a learned data distribution model. 
The algorithm leverages the model to arrange the input records into mutually exclusive, monotonic, and equi-depth partitions that, once sorted, can simply be concatenated to form the output. 
This method completely eliminates the need for multi-way file merging, which is typically used in external sorting. 

We present thorough benchmarks for uniform and skewed datasets in various storage media, where we measure the sorting rates, size scalability, and energy efficiency of \algo and other sorting algorithms. 
We observe that \algo has up to $1.65\times$ higher sorting rates than the next-best external sort (Nsort) on SSD drives and $5.31\times$ higher than the GNU coreutils’ sort utility on Intel\R Optane\TM non-volatile memory.
In addition, \algo supersedes the current winner of the SortBenchmark for the most energy-efficient external string sorting algorithm by an impressive margin of 41\%.
These results reinforce the premise that novel learning-enhanced algorithms can provide remarkable performance benefits over traditional ones. 
\end{abstract}

\maketitle

\section{Introduction}
External sorting is at the core of many operations in large-scale database systems.
In relational databases, it is most obviously used in \texttt{ORDER\_BY}, and \texttt{GROUP\_BY} queries for large result sets\cite{mysql_orderby,postgres_orderby}.
However, it is also part of internal operations, such as building indexes\cite{mysql_bldidx,postgres_bldidx}, sort-merge joins\cite{blas77}, duplicate removal, as well as sharding and record clustering\cite{postgres_cluster}.
External sorting is also a fundamental piece of search engines and other document retrieval systems that use inverted files to map records to their locations on disk\cite{google_inverted_idx}. 
Moreover, it is used in constraint-based systems that work with limited memory and have to rely on persistent storage for their tasks\cite{pang93}.
Hence, improving the speed of external sorting can have a profound impact on large-scale data processing systems. 

The major bottleneck of external sorting is the latency associated with disk accesses, which are an order of magnitude slower than DRAM memory. 
Unlike in-memory sorting, these algorithms need to work together with the operating system and the file system to efficiently utilize system resources and minimize disk I/O.
The majority of external sorting algorithms\cite{kioxia_sort,tencent_sort,nad_sort,unix_sort} that have appeared in leading sorting benchmarks follow the paradigm of \textit{External Mergesort}\cite{clrs}.
Conceptually, these methods first divide the input into small files such that they are small enough to fit entirely in memory. 
Then, these intermediate files are sorted separately and in parallel using Quicksort or alternative in-memory sorts. 
Finally, the sorted files are merged using a heap containing the smallest key from each of them. 
The algorithm continuously appends records to the output file by popping the root of the heap. 

In contrast, this paper describes a parallel external sorting algorithm (\algo), which uses an innovative paradigm based on learned data distribution models.
Unlike External Mergesort, it uses a partition-and-concatenate approach.
\algo first builds a model that estimates the input distribution based on a sample of keys. 
Then, it uses this model to arrange the input records in mutually exclusive, monotonic, and equi-depth partitions. 
In this way, the model eliminates the need for merging, and \algo only needs to sequentially concatenate the partitions, which is much faster than a multi-way file merge. 
While conceptually similar to Radix Sort, model-based partitioning is less sensitive to skew and produces balanced loads for the worker threads.

For in-memory sorting, \algo uses the LearnedSort algorithm\cite{learned_sort}, which has shown high performance in numerous data sets with various distributions. 
Note that simply plugging in LearnedSort as the internal routine in the External Mergesort framework will not considerably affect performance since external sorting is disk I/O-bound.
The algorithm spends most of the execution time performing read/write operations from and to disk; therefore, only I/O-related improvements can make a significant impact on the algorithm's throughput. 
Our design goals for \algo were to have lock-free concurrency, load balance, and optimized file I/O. 

Currently, \algo supports sorting of ASCII strings as those are still the most common data types and multiple systems only support ASCII characters natively\cite{dns_ascii, php_ascii, mysql_ascii}. 
We do not yet support Unicode as it requires special handling functions\footnote{For example, there is no definition for how to sort Unicode strings because the lexical order is language-dependent\cite{unicode_collation}(e.g., \"{o}<z in German, but \"{o}>z in Swedish).}, though we consider extending \algo to Unicode in future work. 

To evaluate \algo, we measure its sorting rates in various machines, data distributions, and input sizes. 
We use desktop and server-grade machines containing multiple storage technologies (i.e., HDD, SSD, and PMem) with varying I/O bandwidths and interfaces (SATA, PCIe, and DDR-T).
We also tested \algo's performance for terabyte-scale uniform and skewed data sets that are up to $40\times$ larger than memory capacity. 
Our implementation showed notable performance gains over existing external sorting algorithms - up to $1.65\times$ higher sorting rates than the next-best external sort (Nsort) on SSD drives and $5.31\times$ than the GNU coreutils' \texttt{sort} utility on Intel\R Optane\TM non-volatile memory. 
These results were consistent among the spectrum of our evaluations, reinforcing the premise that novel learning-enhanced algorithms can provide extraordinary performance benefits to traditional algorithms.

Finally, we also evaluated \algo using the SortBenchmark's JouleSort metric, which assesses the energy efficiency of external string sorting algorithms\cite{joulesort}. 
As of this publication, \algo outperforms the current winner by 41\%; an arguably impressive result that we have officially submitted to the SortBenchmark committee to be considered for the next official ranking. 

In the remainder of this paper we:
\begin{enumerate}
    \item Introduce a new parallel external sorting algorithm for ASCII records that uses learned models,
    \item Describe our method for representing ASCII strings onto a numerical space for modeling their distribution and empirical CDF, 
    \item Analyze the complexity of this new sorting algorithm, and
    \item Present benchmarks that showcase the performance of \algo in various hardware compositions
    \item Discuss the strengths of \algo with respect to other external sorting algorithms. 
\end{enumerate}

The following section gives some background on existing approaches for external sorting, their variations, and their differences.
\section{Background \& Existing approaches}

External sorting algorithms are typically implemented in two phases: 1) The \textit{Run Creation} phase and 2) the \textit{Merge} phase \cite{mysql_extmergesort,postgres_extmergesort,sqlite_extmergesort, unix_sort, kioxia_sort, tencent_sort, nad_sort}. 
In the Run Creation phase, the sorting algorithm divides the input file into several chunks, which are also called \textit{runs}. 
This is usually done in a parallel fashion to speed up the file reading performance.
The objective of this phase is to create intermediate files containing runs that are small enough to fit entirely in memory. 
At the end of the first phase, each intermediate file is sorted using an in-memory sorting algorithm, most commonly Quicksort\cite{unix_sort,mysql_extmergesort,postgres_extmergesort,kioxia_sort}.
Several runs can be sorted in parallel depending on available memory since there are no dependencies between them.  

In the second phase, the sorted runs are merged onto the output file similar to the in-memory Mergesort algorithm.
This algorithmic template is called External Mergesort and is the one used by the \texttt{sort} utility in GNU Coreutils\cite{unix_sort}.

\subsection{Variations in the implementation of the External Mergesort }
There are many variations of External Mergesort that aim to increase the algorithm's performance, such as the choice of the in-memory sorting routine, I/O operations scheduling, and merge strategies.
Table~\ref{tab:summary_of_algos} summarizes the approaches discussed below.

\begin{table*}[t]
\caption{Summary of the characteristics of some external sorting algorithms. }
\label{tab:summary_of_algos}
\centering
\small
\begingroup
\begin{tabular}{|l|l|l|l|l|}
\hline
\rowcolor{lightgray}
\textbf{Algorithm's name}   & \textbf{Framework type}  & \textbf{In-memory sort} & \textbf{Merging strategy} & \textbf{Known for} \\ \hline
Unix sort          & External Merge-Sort      & Mergesort               & Binary heap   & Default filesort in Linux \\ \hline
Nsort              & Unknown                  & Unknown                 & Unknown   & Many winning SortBenchmark entries \\ \hline
Kioxia Sort        & External Merge-Sort      & Quicksort               & Hierarchical    k-way  & Current winner of JouleSort \\\hline
MySQL filesort     & External Merge-Sort      & Introsort               & Hierarchical k-way   & Variation of Ext MS used in MySQL\\ \hline
Postgres tuplesort & External Merge-Sort      & Quicksort               & K-way & Variation of Ext MS used in Postgres \\ \hline
SQLite vdbesort    & External Merge-Sort      & Mergesort               & Hierarchical k-way   & Variation of Ext MS used in SQLite \\ \hline
\algo               & Partition \& Concatenate & LearnedSort             & N/A  & The algorithm described in this paper \\ \hline
\end{tabular}
\endgroup
\end{table*}

\paragraph{In-memory sorting variations}
The most trivial variation is the choice of the in-memory sorting routine used to sort the created runs.
For instance, some algorithms use Timsort, a hybrid of the internal Mergesort algorithm and Insertion Sort.
The rationale is that Timsort leverages the naturally occurring patterns in the input data and has better performance than Quicksort for real-world datasets\cite{timsort}.
Timsort first finds pre-sorted sequences of elements in the input and merges them to create a total order. 
When it cannot find large enough sequences (typically greater than 32 elements), it employs Insertion Sort to form sorted sequences from consecutive out-of-order elements. 
Timsort is used as an internal sorting routine by algorithms like NADSort\cite{nad_sort} and SparkSort\cite{spark_sort}. 

\paragraph{I/O scheduling variations}
External sorting algorithms are typically disk I/O-bound, so compute-related improvements, like the choice of the internal sorting routine, do not help much with the more significant issue.
The most successful external sorting strategies focus on 1) minimizing the number of I/O operations and 2) hiding I/O latencies by overlapping them with as much computation as possible.
For example, KioxiaSort, one of the current winners of the SortBenchmark\cite{kioxia_sort}, addresses this issue by adopting a pipeline approach.
Using mutual exclusions, it overlaps the \textit{read} operations of one-third of the threads with the \textit{sort} operations of the second third of the threads and \textit{write} operations of the final third of the threads\cite{kioxia_sort}. 
A similar strategy is also implemented by FuxiSort\cite{fuxi_sort} and DEMSort\cite{dem_sort}, previous winners of the same benchmark. 
Besides overlapping, these algorithms also reduce the amount of I/O by generating smaller intermediate files that do not contain the record payloads. 
Instead, the algorithm only writes out the record keys and indices from the original input file\cite{kioxia_sort}. 

\paragraph{Merge strategy variations}
On the other hand, there are a few prominent approaches to how the intermediate files are merged in the second phase of the External Mergesort algorithm. 
The first strategy is to perform a multi-way external merge, in which the algorithm maintains a heap of the smallest key from each file and writes the element at the root of the heap to the output. 
However, this method is limited by how many elements can be kept in memory, restricting the number of intermediate files that the algorithm can use and the degree of parallelism of the first phase. 
In addition, element insertions to the heap become expensive due to the large heap size. 
Moreover, this merging routine can only be performed sequentially, thus becoming the algorithm's bottleneck. 

Alternatively, the Merge phase can avoid using a heap by performing hierarchical 2-way external merging (like a tournament tree). 
Not only does this method reduce the memory footprint of this phase, but it also enables the lower level of the external merging to be done in parallel.  
However, this comes at increased disk I/O cost, additional intermediate files, and the need for synchronization primitives (i.e., mutexes), which are costly. 

Therefore, the third and best approach is to have a hybrid merging routine: a hierarchical multi-way merge that combines the benefits of having smaller heaps and parallelism. 
This method has the smaller cost of using only a few additional intermediate files for the merging routine and fewer synchronization variables among the hierarchy levels. 
KioxiaSort, for example, uses two-stage merging, where the first stage performs six 200-way merges in parallel, and the second stage performs a 6-way merge to the output file\cite{kioxia_sort}. 

\subsection{Nsort}
Nsort is an interesting external sorting algorithm that is widely used and displays very high sorting throughput\cite{nsort}.
Nsort is employed as the software layer of multiple SortBenchmark winners like RezSort\cite{rez_sort}, TaichiSort\cite{taichi_sort}, NTOSort\cite{nto_sort}, EcoSort\cite{eco_sort}, FlashSort\cite{flash_sort}, and FAWNsort\cite{fawn_sort}. 
Therefore, it is very intriguing to investigate the successful recipe that has made SortBenchmark winners since 2007. 
The authors mention that it has ``sophisticated buffer management to overlap computation and I/O'' and that ``it pays particular attention to processor cache locality to make the best use of fast microprocessors''\cite{nsort_paper}. 
However, its source code remains proprietary at the time that we write this paper, and we do not possess any further insight into this algorithm.

Even though there exist numerous external sorting algorithms, it is clear that none of them move away from the Run-Merge paradigm. 
In fact, there is very little innovation in the algorithmic space, with most of the efforts being spent on building hardware setups that exploit modern high-speed block storage, processors, and network adapters. 
In the next section, we look at a different sorting approach that completely eliminates the need for a Merge phase due to the benefits of using a learned CDF model.

\section{Sorting larger-than-memory files}

\begin{figure}
    \includegraphics[width=\linewidth]{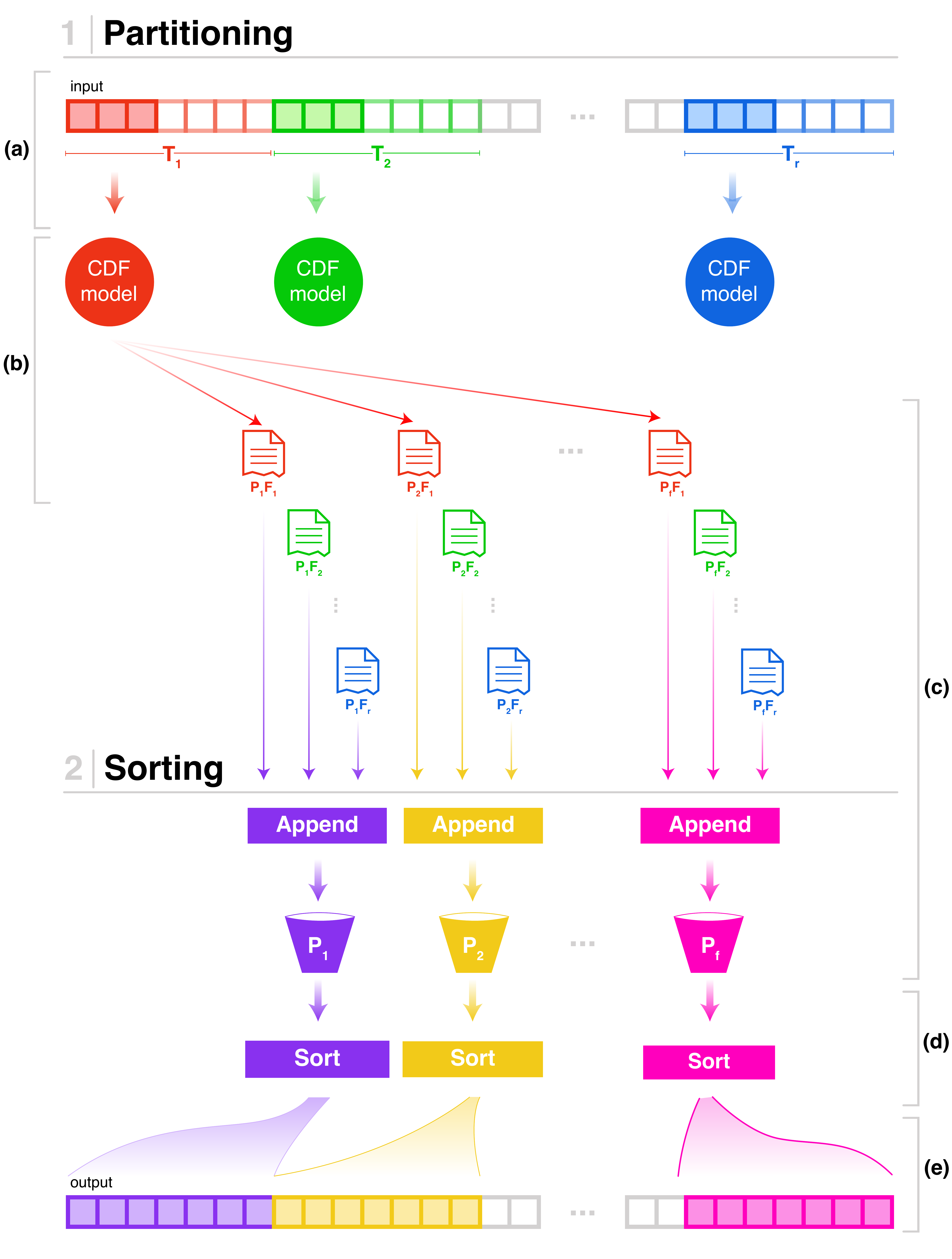}
    \caption{The architecture of \algo. 
    During partitioning, $r$ parallel threads read different stripes of the input file in batches. 
    Each record is placed into $f$ thread-local partition fragments, then flushed out to intermediary files. 
    Partitions are composed of the contents of the corresponding fragment files. 
    They are sorted in memory using LearnedSort\cite{learned_sort} and written directly to the corresponding offset in the output file. }
    \label{fig:algo}
\end{figure}

\algo is a learning-enhanced, data distribution-based, external sorting algorithm that sorts ASCII datasets by leveraging small, highly accurate, and fast ML models. 
The algorithm combines various techniques to achieve high sorting rates, such as sample-based distribution learning, numerical embedding for ASCII keys, and parallel, buffered and lock-free file I/O. 
The design of \algo incorporates I/O optimizations that avoid bottlenecks while maximizing utilization of the memory and available hardware parallelism. 
The central idea of \algo is organizing input records in mutually exclusive, monotonic, and equi-depth partitions that, once sorted, can simply be concatenated to form the output file. 
In order to perform that task fast and with good accuracy, \algo uses a CDF model, which approximates the empirical distribution function of the input dataset. 
Using this model, the algorithm can infer the rank of each record in the input file by simply looking at its key and not making any comparisons with the other records. 
This approach provides better partition size uniformity than a radix-based partitioning scheme, which is sensitive to skew. 

A diagram of \algo is shown in Figure \ref{fig:algo}, and the sections below describe the sorting procedure in detail.

\subsection{Model training}
The CDF model is arguably the central component of \algo. 
The model is trained on a small sample from the input data ($X$) and uses a Recursive Model Index (RMI) architecture, as described in \cite{learned_sort}, and \cite{rmi}. 
The sample is picked uniformly at random from the first batch read by thread $T_0$. 
Its size must be large enough to enable the model to learn well the boundaries of the partitions while not being so large that it makes the model training an expensive operation. 
In \cite{learned_sort} we have observed that the CDF model reaches an adequate accuracy for the partitioning task quickly. 
Hence this approach works very well even with small samples. 
In our implementation of \algo, we empirically observed that a sample of 1\% of the data satisfies these constraints.

On the other hand, the RMI structure is an acyclic graph arrangement of linear regression models that acts as an expert system. 
Starting from the root, the record key from the input ($x\in X$) traverses internal nodes, which recursively pick another linear model that will be more ``specialized'' for a subset of values where the current key belongs. 
When the key reaches the leaf nodes, the output is a predicted CDF value $y \in [0, 1]$ that corresponds to $y = P(X \le x)$, or the percentile rank of the record among the estimated population.

This CDF model combines the benefits of having good accuracy with a fast training and inference time. 
In some sense, the model acts like an order-preserving hash function or a radix partitioner. 
However, unlike these, the CDF model absorbs input skew much better and produces uniformly-sized partitions, which is critical for load balancing the sorter threads\cite{learned_sort}. 
This is because the training procedure assigns high-density domain areas to more nodes in the RMI, hence spreading out the skew onto more evenly-distributed buckets, ideally acting as equi-depth histogram bins. 

It is important to note that we are interested in modeling the \textit{empirical} distribution, i.e., the distribution of the \textit{observed} data, rather than the theoretical one. 
The difference is that the observed sample does not follow a smooth distribution that could be modeled with just a few linear models. 
The data will behave like a step function at a fine scale, with more structure, noise, and irregularities. 
That is why it is necessary to use a more complex architecture to capture the empirical distribution's subtle characteristics. 

``The Case for Learned Index Structures''\cite{rmi} provides a detailed explanation of the architecture and the CDF model's training algorithm, inference, and further analysis.

\subsection{Input processing}
After the training has been completed, the algorithm spawns $r$ threads, each responsible for reading a specific, non-overlapping range of records from the input file (Fig~\ref{fig:algo}a).
In our implementation, we used OpenMP threads due to (1) their better portability than pthreads, (2) thread pooling, which minimizes thread setup and teardown costs, (3) automatic scheduling, and (4) CPU affinity. 
Each thread reads the input records (key and payload) in batches. 
This technique aims to utilize as much sequential I/O as possible to take advantage of the higher reading speeds than random access I/O. 
Nevertheless, in any sorting procedure, record shuffling has unpredictable patterns that make it almost impossible to perform sequential reads and writes. 
In another attempt to speed up reading times, we could use asynchronous threads (or co-routines) to read ahead of the next batch while the reader threads are processing the current batch. 
However, we observed that this extension did not improve the performance due to the synchronization overhead and the diminished parallelism. 
In addition, we also did not observe significant performance improvements from using direct I/O, therefore the algorithm relies on     
\texttt{libc}'s buffered file utilities. 

\subsection{Partitioning}
Next, the algorithm processes the keys in the batch through the trained CDF model, which predicts the ranks of the records among all records in the input (Fig~\ref{fig:algo}b). 
Each reader thread $T_i$ has their own read-only copy of the model and maintains a set of $f$ thread-local partition fragments that act like equi-depth histogram bins (i.e., $P_1F_i, P_2F_i, ..., P_fF_i$).
Using the predictions from the model, the threads place the records in respective fragments.
For the $j^\text{th}$ fragment $P_jF_i$ in the $i^\textsuperscript{th}$ reader thread, the algorithm maintains the following monotonicity invariant:  

\begin{equation}
    \label{eq:invariant}
    \begin{split}
        \{x \leq y \mid\forall x\in P_jF_i, \forall y\in P_{j+1}F_i, \\
        i\in[1, f], j\in[1,f-1]\}
    \end{split}
\end{equation}

Once the first batch processing is done, the threads flush their partition fragments to temporary files. 
For future batches, the contents of the fragments are simply appended to the existing and corresponding temporary files.
Note that each reader thread maintains its thread-local partition fragments instead of directly appending them to a big partition file. 
This avoids using mutexes and locks, which are costly. 
On the other hand, the algorithm only uses pointers to the records during this process to reduce memory copying and moving operations, especially for long strings. 
Finally, since the reader threads have mutually disjoint working sets, we can perform file I/O with the non-locking versions of the read and write functions (i.e., \texttt{fread\_unlocked()} and \texttt{fwrite\_unlocked()}). 
These omit the file pointer's lock check and are faster.

We compared this partitioning approach with a radix-based one on large skewed datasets and observed that the CDF model provides better uniformity. 
Radix-based partitioning looks at the most significant bytes in the key and converts them to indices of the partition files. 
This approach is similar to building an equi-width histogram, where each bin corresponds to a fixed interval of the key domain that has equal width to the other bins. 
In contrast, the model-based partitioning used in \algo produces equi-depth partitions with variable-sized key intervals but evenly-sized bins. 
Based on our experiments, our approach reduced the partition size variance by 23\% compared to the radix approach. 
This will provide a better load balance for the sorter threads that will process the partition files in the next stage. 

\subsection{Sorting}
After the entire input file has been processed, \algo spawns $s$ threads, each responsible for sorting and flushing one partition.
The number $s$ is calculated as the maximum number of partitions that can be in memory simultaneously. 
Note that the number of partitions is chosen such that no single partition exceeds the memory capacity. 
Furthermore, this allows for several partitions to be in memory simultaneously, thus enabling parallel sorting. 

The sorter threads read all the fragment files that belong to the logical partition that they are assigned to and append their records into a single large buffer (Fig~\ref{fig:algo}c). 

Then the algorithm calls LearnedSort as an in-memory sorting routine (Fig~\ref{fig:algo}d). 
This is an excellent choice because it has the highest in-memory sorting rates compared to many modern sorting algorithms\cite{duplicates_ls}. 
It implements various optimization related to CPU cache utilization, handling skewed inputs, and avoiding performance degradation on high-duplicate datasets by using an early termination strategy\cite{duplicates_ls}.

\subsection{Writing to output}
Finally, after the partition contents have been sorted, they are concatenated sequentially with the neighboring partitions' records to form a single continuous output file (Fig~\ref{fig:algo}e). 
Each sorter thread maintains an open descriptor for the output file, and, for each partition $P_i$ that it will flush, it seeks to the offset location pre-calculated as the sum of the sizes of the partitions $P_1..P_{i-1}$:
$$write\_offset = \sum_{k=1}^{i-1}len(P_k)$$

Since the sorted buffer contains only keys and pointers to the records, it is impossible to flush the entire buffer in one sequential call. 
Therefore, the thread first coalesces the records in batches by dereferencing the pointers in sorted order.
Then, it performs a buffered, sequential write of the coalesced buffer (typically 100KB), thus optimizing the write performance. 

\section{Encoding ASCII records}
In order to learn the distribution of the input, the ASCII records have to be projected onto a numerical space on which the CDF model can use linear regression to train its individual nodes. 
Therefore, the algorithm operates on the key's numerical encoding and a pointer to the record while in memory. 

Assuming the keys have a fixed size, the encodings are calculated using the binary values of each character in the key represented as base-95 numbers since printable ASCII characters have codes between 32 and 127. 
The unprintable ASCII characters (0-31) are control codes intended to provide meta-information for peripherals (e.g., printers), so they are not of interest in sorting. 
Therefore, encoding of a character in position $x_i$ of the key of length $l$ ($1\leq i \leq l$) is $(\textsc{ASCII}(x_{i}) - 32) \times 95^{l-i}$. 
Then, the numerical encoding of the entire key is:
$$\sum^l_{i=1}\left((\textsc{ASCII}(x_i) - 32) \times 95^{l-i}\right)$$

If the input contains variable-sized keys, then $l$ is set to the maximum length observed ($l=\max(len(x_1),..., len(x_n)$), and $\textsc{ASCII}(x_i)=0$ for $i\geq len(x)$.

With a 64-bit primitive type, we can encode up to the ninth byte of the key, since $\log_2\sum^l_{i=1}95^{l-i+1}$ is $59.14$ for $l=9$ and $65.71$ for $l=10$. 
If the record keys are longer than nine bytes, this encoding scheme will not be able to capture the rest of the bytes numerically. 
This only violates the invariant in Eq.~\ref{eq:invariant} if we use more than $6.37\cdot10^{17}$ partitions and the keys are at least 10 bytes, meaning that we are sorting inputs of at least 6369.5 Petabytes. 
In all other cases, this will not affect the correctness of the sorting algorithm. 
This is because the in-memory LearnedSort routine has a touch-up step that performs last-mile sorting on the rest of the key using pair-wise key comparisons with the \texttt{strncmp()} function.
In addition, this touch-up step also covers prediction inaccuracies from the CDF model. 
This step uses Insertion Sort, which works in almost linear time for nearly-sorted arrays\cite{learned_sort}. 

\section{Implementation}
The complete pseudocode of the \algo algorithm is shown in Algorithm~\ref{algo}. 
The algorithm's inputs are $I$-  the input file name, $O$ - the output file name,  $B$ - the batch size, $f$ - the number of partitions/fragments, $r$ - the number of processors, and $M$ - the available memory in the system.

\textbf{In line 1}, the algorithm creates a sparse output file that occupies precisely as many bytes as the input file $I$. 

\textbf{In line 2}, the algorithm calls a model training function that samples the input file and returns a trained CDF model (i.e., $F_X(x)$).

\textbf{In line 3}, the algorithm initializes a 2D matrix ($r\times f$) of dynamically-sized arrays (vectors). 
This matrix will store the records read in memory organized by the reader thread on the first dimension and the partition fragment on the second one. 

\textbf{In line 4}, the algorithm initializes a 2D matrix ($r\times f$) of temporary files that correspond to the records collected in $D$.

\textbf{In line 5}, the algorithm creates a size vector that will keep track of the number of records assigned to each partition. 

\textbf{In lines 6-20}, the algorithm starts by spawning $r$ reader threads, each opening the input file at different offsets ($o$) for parallel reading.
Note that each thread will be responsible for reading $|I|/r$ records from the input file. 
Each reader thread $i$ reads the records that it is responsible for in batches ($\textbf{C}$) of size $B$ while keeping track of the number of bytes read so far ($b$). 
Then, for every record in each batch, it uses the $F_X(x)$ predictor to estimate the rank of the key ($p$) and place the record in the predicted fragment ($\textbf{D}_{i,p}$).
For this call, the algorithm encodes the ASCII keys using the method described in the section above. 
Once a record has been placed in a fragment, the corresponding partition's size (i.e., $S_{p}$) is incremented. 

\textbf{In line 21}, the algorithm transitions to the second stage, where it starts by calculating $s$ - the number of sorter threads to be used. 
Recall that the value of $s$ is the maximum number of consecutive partitions that can be completely held in memory simultaneously. 

\textbf{In lines 22-31}, the algorithm spawns $s$ working threads, which will process all $f$ partitions. 
Therefore, each sorter thread will be responsible for $f/s$ partition files. 
Here the algorithm also initializes an in-memory buffer that will contain all the records belonging to these partitions that are, up to now, separated onto $r$ temporary fragment files from earlier. 
Each sorter thread $i$ reads the content of the fragments belonging to the partition that they are processing ($\textbf{F}_{j,i}$) and gathers their contents onto a single in-memory buffer for sorting ($\textbf{P}$). 
After each fragment is fully read into memory, its temporary file is closed, immediately signaling the OS to remove the file and free up memory. 
Then, the sorter invokes the LearnedSort internal sorting routine to sort the partition contents. 
This call will internally re-encode the keys of each record to a numerical representation and use it to sort the keys as described in \cite{duplicates_ls}. 
Finally, the sorter will flush the entire partition to its respective offset $o$ of the output file, such that $o$ is the sum of sizes of all partitions $[0, i-1]$.
This parallel writing routine is simply an in-order concatenation of the partitions. 
Once every sorter thread has processed its corresponding partitions, the algorithm exits, and the output file is fully-populated with sorted records. 

\SetKwComment{Comment}{/* }{ */}
\begin{algorithm}[hbt!]
\DontPrintSemicolon 
\caption{The pseudocode of \algo}\label{algo}
\KwInput{$I$ - Input file name}
\KwInput{$O$ - Output file name}
\KwInput{$M$ - available memory}
\KwInput{$r$ - number of processors}
\KwInput{$f$ - number of partitions}
\KwInput{$B$ - reading batch size}
    $out \gets fcreate(O, |I|)$
    
    $F_X \gets train\_CDF(I)$
    
    $\textbf{D}_{r\times f} \gets {vector()}$
    
    $\textbf{F}_{r\times f} \gets {tmpfile()}$
    
    $\textbf{S}_{1\times f} \gets \{0\}$
    
    \For(\textbf{in parallel}){$i \in [0, r)$}{
        
        $o\gets i \cdot |I| / r$
        
        $in \gets fopenat(I, ``rb", o)$
        
        $b \gets 0$
    
        \While{$b < |I| / r$}{
        
            $\textbf{C}\gets fread(in, B)$
            
            $b \gets b + B$
            
            \For{$x \in \textbf{C}$}{
                
                $p \gets F_X(encode(x))$
                
                $\textbf{D}_{i,p} \gets \textbf{D}_{i,p} \cup \{x\}$
                
                $\textbf{S}_{p} \gets \textbf{S}_{p} + 1$
            }
            
            \For{$j \in [0,f)$}{
               
               $fwrite(\textbf{F}_{i,j}, \textbf{D}_{i,j})$
               
               $clear(\textbf{D}_{i,j})$
            }
            
        }
        
        $fclose(in)$
    }
    
    $s\gets\{x\mid\sum^x_{i=0}\textbf{S}_{i}\le M < \sum^{x+1}_{i=0}\textbf{S}_{i}\}$
    
    \For(\textbf{in $s$ parallel threads}){$i \in [0, f)$}{
    
        $\textbf{P} \gets \{\} $
        
        \For{$j \in [0,r)$}{
         
            $\textbf{P} \gets \textbf{P} \cup \{fread(\textbf{F}_{j,i})\}$
            
            $fclose(\textbf{F}_{j,i})$
         
        }
         
        $LearnedSort(\textbf{P})$
        
        $o \gets \sum^{i}_{l=0}S_l$
        
        $out \gets fopenat(O, ``wb", o)$
        
        $fwrite(out, \textbf{P})$
        
        $fclose(out)$
    }
\end{algorithm}

\section{Computation complexity}
This section provides a Work-Span analysis of \algo with respect to the input size ($n$).
The work-span framework in a Parallel Random Access Machine (PRAM)\cite{jaja} model is similar to the complexity analysis of sequential algorithms. 
However, it differs in that the computation is treated as a directed task dependency graph.
The term \textit{work} refers to the total number of operations executed by the algorithm, which represents a lower bound of the worst-case sequential computation complexity. 
On the other hand, \textit{span} represents the maximum number of sequential operations (i.e., the critical path) that cannot otherwise be shortened due to data dependencies while assuming unbounded parallelism\cite{clrs, jaja}.

The algorithm starts by allocating enough space in the disk for the output file, equal to the input size. 
If the underlying filesystem supports sparse files, this is an $O(1)$ operation (i.e., XFS, EXT4, NTFS, APFS.). 
Otherwise, this operation is linear w.r.t. the input size. 
Next, \algo takes a small sample from the input and trains the CDF model. 
Since the sample size is capped at 10M for inputs larger than 1B, this is a constant operation asymptotically. 

Then, the algorithm switches from sequential to parallel mode during the input reading phase.
The algorithm uses $r$ threads that read equal-sized ranges from the input (i.e., each is responsible for $n/r$ records) and shuffle them into the corresponding partition fragments. 
Since the CDF prediction is made in constant time using the RMI model, the partitioning phase has $O(n/r)$ span and $O(n)$ work. 
In this case, we assume that the storage device has random read/write access and that the seek time is constant. 
Note that this would not be the case in spinning disks; for example, the work for the parallel read operation would increase to $O(n(r-1)/2)$ due to the non-constant seek time. 

After the records in each reader thread's batch have been placed into the predicted partition fragments, fragments across threads belonging to the same partition are appended to the corresponding partition file. 
There are $f$ partition files in total, and the operation is done in $s$ parallel threads, where $s$ is the maximum number of partitions that can be fully kept in memory at the same time. 
Each partition file will contain, on average, $n/f$ records, and each of the $s$ sorter threads will be responsible for $f/s$ partition files on average. 
Therefore, the span of this operation is $O(n/s)$, and work is $O(n)$. 

Once each partition file has been completed, the sorter threads invoke LearnedSort, whose complexity is linear with respect to the input size in the average case\cite{duplicates_ls,learned_sort}. 
Therefore, the span of this operation is $O(n/s)$ and work is $O(n)$. 

Finally, each sorter thread flushes the sorted partition contents to a specific offset of the output file. 
Again, assuming that the storage device has random write access, the seek time would be constant, and the span of this operation would be $O(n/s)$ and work $O(n)$. 
Otherwise, the seek time for each sorter would be $O(n(f-1)/(2s))$. 

In total, the span is $O(n/r + 3n/s)$ and work is $O(n)$. 
\begin{table*}[!ht]
\caption{Key statistics of the machines used for performance evaluation of \algo.}
\label{tab:machines}
\centering
\small
\setlength{\tabcolsep}{3pt}
\setlength{\extrarowheight}{2pt}
\begingroup
\renewcommand{\arraystretch}{1.2} 
\begin{tabular}{|l|l|c|c|l|l|l|c|}
\hline
\rowcolor{lightgray}
\hline
 & \textbf{CPU Model} & \multicolumn{1}{l|}{\textbf{CPU freq}} & \multicolumn{1}{l|}{\textbf{SMT threads}} & \textbf{Disk Model} & \textbf{Disk Type} & \textbf{Interface} & \multicolumn{1}{l|}{\textbf{Memory}} \\ \hline
\multirow{3}{*}{\textbf{McGraw}} & \multirow{3}{*}{Intel\R Xeon\R Gold 6230} & \multirow{3}{*}{2.1GHz} & \multirow{3}{*}{40} & Intel\R SSDSC2KB03 & SSD & SATA III & \multirow{3}{*}{256~GB} \\ \cline{5-7}
 &  &  &  & Intel\R SSDPE2KX010T8 & SSD & NVMe &  \\ \cline{5-7}
 &  &  &  & Seagate ST4000NM0115-1YZ & HDD & SATA III &  \\ \cline{5-7}
 &  &  &  & Intel\R Optane\TM 100 Series & PMem & DDR-T &  \\ \hline
\textbf{Aurora} & Intel\R Core\TM i5-12600K & 3.7GHz & 16 (12P + 4E) & 4 $\times$ WD\_BLACK SN850 (2TB) & SSD & M.2 NVMe & 32GB \\ \hline
\end{tabular}
\endgroup
\end{table*}

\section{Benchmarks}

In this section, we will evaluate the performance of \algo against other popular external sorts on ASCII datasets and various machine configurations.
First, we describe the benchmark data and the hardware used for the experiments. 

\subsection{Experimental Setup}
\paragraph{Baselines} 
For the following experiments, we will compare the sorting rates of \algo with GNU coreutils' \texttt{sort} utility (a.k.a Unix sort)\cite{unix_sort} and Nsort\cite{nsort}. 
Unix sort is an External Mergesort and it represents the most popular external sorting paradigm (i.e., Run Creation \& Merging) that is used by numerous DBMS like MySQL\cite{mysql_extmergesort}, Postgres\cite{postgres_extmergesort}, and SQLite\cite{sqlite_extmergesort}. 
On the other hand, Nsort is the highest performing external ASCII sort that we are aware of and is publicly available. 
Note that, based on the results displayed on the SortBechmark website, KioxiaSort is also a very competitive algorithm.
However, despite multiple attempts, we have been unable to obtain the code from its authors. 

\paragraph{Data} 
We perform the evaluations on single-file datasets containing printable ASCII strings of fixed key and payload sizes.
The key size is 10 bytes, whereas the payload is 90 bytes, hence allowing for $95^{10}$ possible key values. 
We show results on two datasets: skewed and not skewed, which are generated using the \texttt{gensort} utility\cite{gensort}. 
For the uniformly-distributed input, gensort produces each key character independently and with equal probability.
Whereas for the skewed case, it first generates non-skewed records, then modifies the dataset to inject some skew. 
For this, it maintains a table of 128 6-byte entries, and for each record with index $rec\_idx$, it substitutes the most significant bytes of the key with the table entry at index $table\_idx = \log_2(rec\_idx) \mod 128$. 

\paragraph{Machines} 
We use two different Linux machines, one server-grade, and one desktop-grade machine. 
The server-grade machine, McGraw, is connected to different types of storage devices, i.e., HDD, SATA SSD, NVMe SSD, and Intel\R Optane\TM persistent memory. 
This makes it interesting to compare the performance of the sorting algorithms on disks of various bandwidths and characteristics. 
This machine contains an Intel\R Xeon\R Gold 6230 processor with massive parallelism (40 threads) but not the highest clock speeds. 
On the other hand, Aurora, the desktop machine, contains an Intel\R Core\TM i5-12600K processor, which can achieve much higher clock speeds. 
This processor has a hybrid architecture with six high-performing cores and four power-efficient ones. 
Aurora contains 4 2TB NVMe SSDs, each with an M.2 form factor directly connecting to the motherboard slots. 
Given that this machine has 32~GB of RAM and 8TB of disk space, it allows us to perform scalability experiments, where we can measure the performance of the sorting algorithms on datasets up to 40x the memory capacity. 
Table \ref{tab:machines} summarizes key specifications that influence sorting rates.

\paragraph{Methodology} 
We sort the records based on the ASCII binary order, as determined by the \texttt{strncmp(3)} or \texttt{memcmp(3)} functions in the GNU C library\cite{libc}.
It is important to note that all the experiments access disk data only via the filesystem mode, hence leveraging buffered file I/O mechanisms of the OS.
All measurements use the maximum number of available threads in the system: 40 for the experiments in the McGraw machine and 16 for the ones in the Aurora machine. 
For each measurement, (1) we empty the system buffer, (2) record the elapsed time via the Unix \texttt{time} utility which includes the process setup and shutdown times, and (3) verify the sortedness and checksum of the output file using \texttt{valsort}\cite{gensort}. 
We report the mean of five consecutive runs of each algorithm. 

\subsection{Different Storage Types}

\begin{figure*}[ht]
    \includegraphics[width=\linewidth]{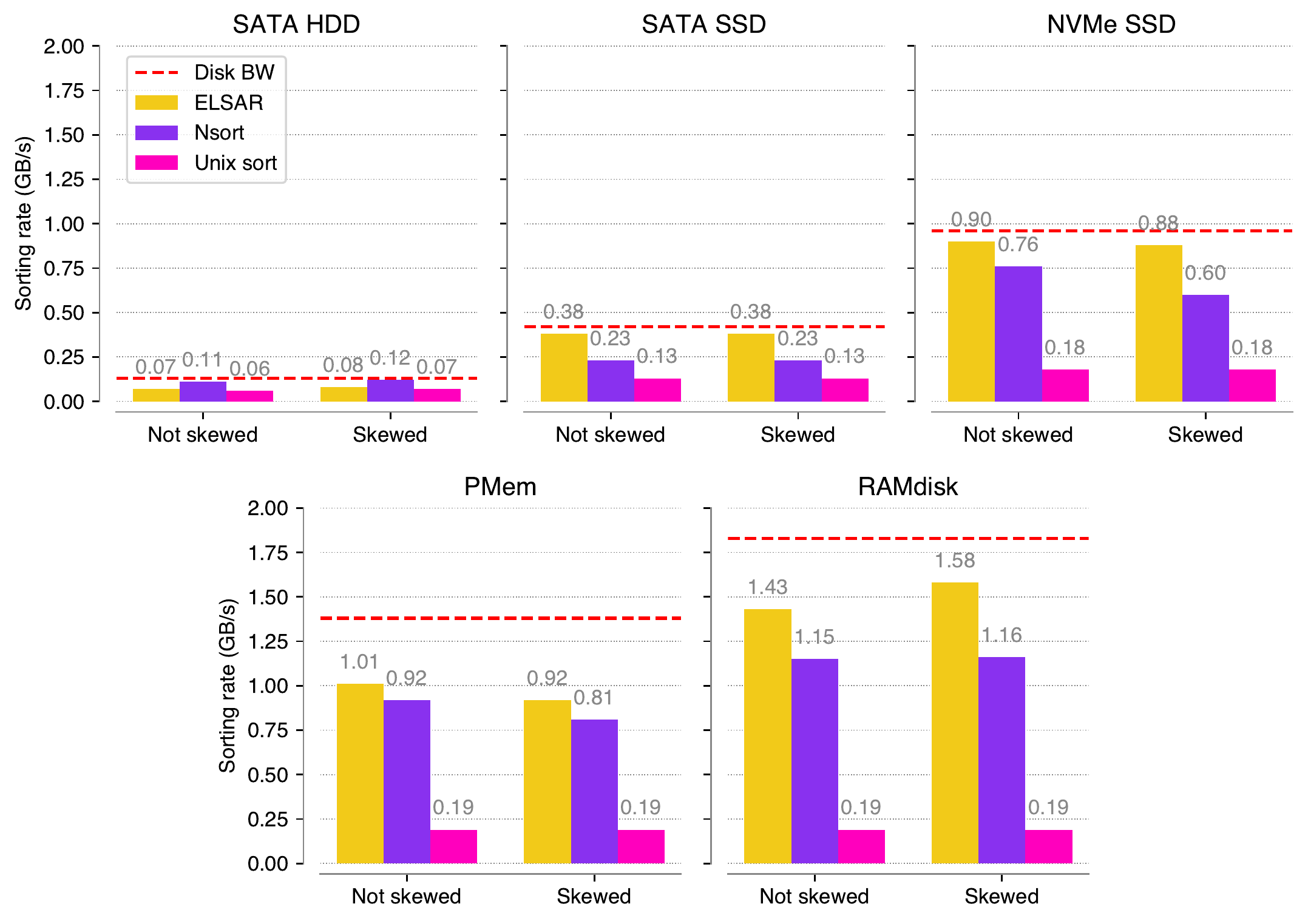}
    \caption{The sorting rates of \algo and other baselines for different storage types belonging to the McGraw machine (higher is better).
    This graph also shows the performance on both uniform and skewed datasets. 
    The reference line (dashed) represents the bandwidth of the disk, measured as the time to fully read the input file and immediately write it out in the same directory.
    All these measurements are performed using the maximum amount of parallelism available on the hardware.}
    \label{fig:storage_types}
\end{figure*}

As we mentioned earlier, the disk I/O speeds heavily influence external sorting algorithms, as they make up most of the execution time. 
Therefore, we start by looking at the sorters' performance on the McGraw machine, which is connected to four different types of storage: SATA HDD, SATA SSD, NVMe\TM SSD, and Intel\R Optane\TM DC Persistent Memory. 
The results are shown in Figure \ref{fig:storage_types}.
We show the disk bandwidth in the dashed red line for each storage device. 
The bandwidth is calculated by reading the input file and immediately writing it back to the same disk using the same number of threads as the sorting algorithms. 
In this section, we only discuss the non-skewed dataset. 

In the case of the spinning hard drive, the bandwidths are very low, and \algo's sorting rate is 70~MB/s. 
This is lower than Nsort's (110~MB/s) and approximately equal to Unix sort (60~MB/s). 
This is a result of \algo heavily depending on disk seeks to maximize the threads' parallelism and minimize their co-dependence. 
Unlike SSDs, the file-seeking operations in hard disks are not constant with respect to the seek offset. 

On the other hand, in the SATA SSD, \algo sorts at 380~MB/s, whereas Nsort at 230~MB/s, which is 39\% lower.
Unix sort's rate is 130~MB/s, which is 66\% lower than \algo. 
\algo's throughput is only 10\% lower than the disk's bandwidth. 

As expected, for the NVMe\TM disk, all three algorithms have higher throughputs.
\algo gets a 2.35$\times$ boost, Nsort $3.35\times$, whereas Unix sort only a $1.41\times$. 
The sorting rate of \algo is 900~MB/s, which is 18\% higher than the next-best algorithm - Nsort (760~MB/s), and, again, only 7\% lower than the disk's bandwidth.

An exciting part of this experiment is the performance of these algorithms on Intel\R Optane\TM DC Persistent Memory (PMem), which is a non-volatile memory technology that enables memory-like performance at storage-like capacity and costs\cite{optane}. 
The Optane\TM devices are connected to the motherboard via the DIMM slots, hence approximating memory bus speeds for reads and writes. 
However, since they are non-volatile media, the technology still does not operate as fast as DRAM speeds, placing it between DRAM and SSD in the performance hierarchy\cite{optane_paper}. 
Nevertheless, \algo still maintains its lead, sorting 13\% faster than on NVMe\TM and outperforming Nsort by 9\%. 
In this case, \algo's gap with respect to the reference line is larger because we have not tailored our code to use the \texttt{libpmem.h} library that contains intrinsics for accessing the Optane\TM drives\cite{libpmem}. 

The final storage type is a RAMdisk, a memory-mounted file system (e.g., ramfs, tmpfs) that simulates a nearly zero-latency disk. 
This is an interesting experiment because we demonstrate how these algorithms behave when they are not disk I/O-bound anymore. 
Storage media will keep getting faster over time, and this evaluation approximates the sorting rates in high-speed disks. 
In this case, \algo's sorting rate jumps to 1.43~GB/s, being 20\% higher than Nsort, about $7.53\times$ higher than Unix sort, and 22\% lower than the RAM's bandwidth. 

\subsection{Skewed datasets}

\begin{figure}
    \includegraphics[width=\linewidth]{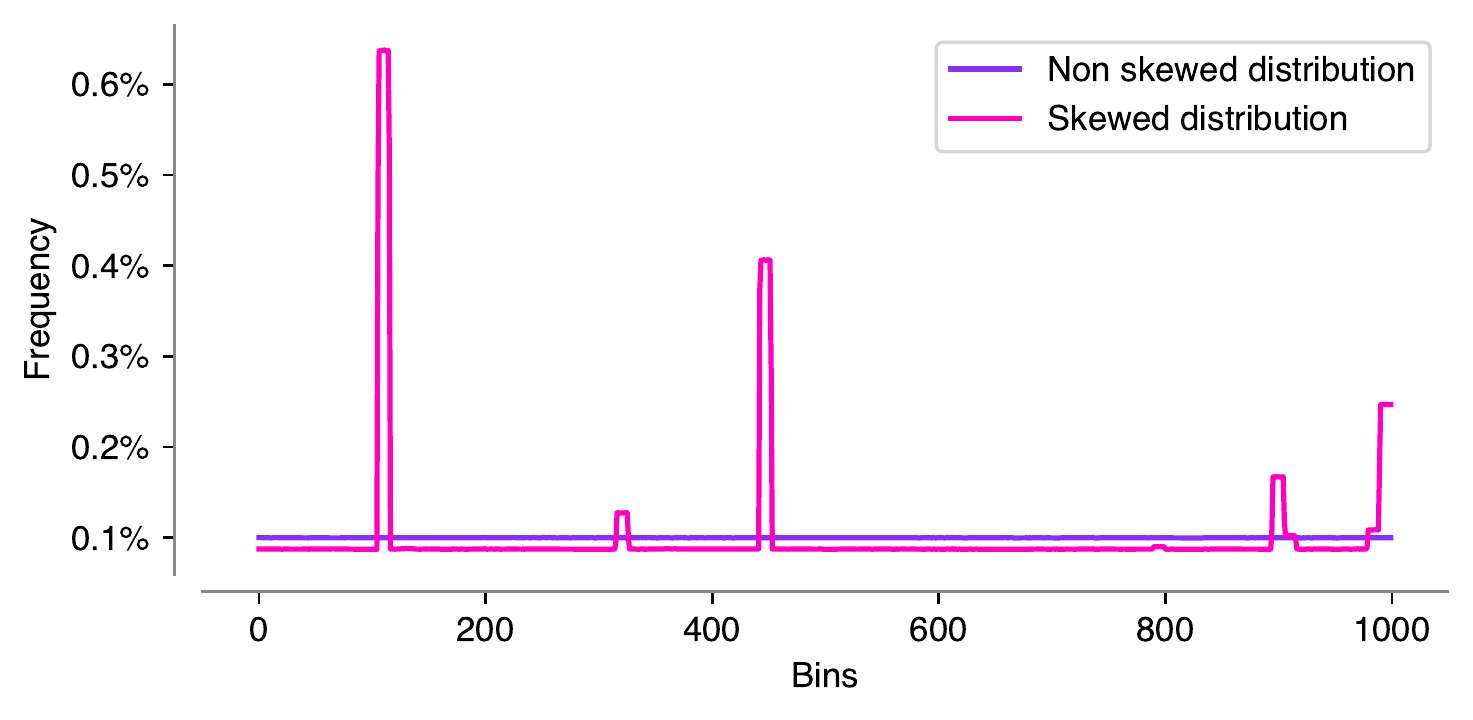}
    \caption{A histogram of the skewed and non-skewed datasets produced from \texttt{gensort}. 
    The histogram contains 1000 bins (x-axis), and the frequency of items is shown on the y-axis, and it is calculated as the ratio of the bin's size to the input size. 
    In the case of the skewed dataset, certain histogram bins can be up to $6\times$ larger than the mean.
    Note that there are smaller ``spikes'' in the bin sizes, but they are not visible due to the scale of the y-axis. 
    The standard deviation of the bin sizes for the non-skewed dataset is 0.14\% of the mean bin size. 
    For the skewed dataset, the standard deviation of the bin sizes is 65.65\% of the mean bin size. 
    }
    \label{fig:hist_skewed_vs_nonskewed} 
\end{figure}

In addition to the performance on different storage types, Figure~\ref{fig:storage_types} also compares the sorting rates on skewed datasets for each disk type. 
The skewed datasets were produced using the \texttt{gensort}\cite{gensort} utility with the \texttt{-s} option. 
This produces various ``spikes'' in the data histogram, increasing the variance in the size of the histogram bins (see Fig.~\ref{fig:hist_skewed_vs_nonskewed}). 
In the case of the skewed dataset, the standard deviation of the bin sizes increases from 0.14\% of the mean bin size to 65.65\% of the mean bin size ($\sim468\times$ higher). 

Figure~\ref{fig:storage_types} shows that, while the sorting rates are slightly affected by the skew, there is no severe performance degradation and \algo is able to absorb the skew. 
The reason for this is twofold: 1) The CDF model predictions create equi-depth partitions, as described in \cite{learned_sort}, since the partitions do not have fixed value ranges, and 2) even if some partitions are larger than other ones, and the sorting time would take longer, OpenMP's dynamic thread scheduling manages to balance the load by assigning the under-worked threads more partitions to work on. 

The results show that, for storage devices, \algo's rate drops by an average of 3\%, and Nsort's by 11\%. 
The performance on hard disks improves by 5\% for both algorithms due to reduced random seeks. 
In the case of the RAM disk, since the algorithms are not disk-bound anymore, \algo's sorting rate is 10\% higher than the non-skewed case. 
This is because the input, output, and temporary files are all in memory, and repeated access to the same few files leverages page cache locality. 

\subsection{Data Scalability}

\begin{figure}
    \includegraphics[width=\linewidth]{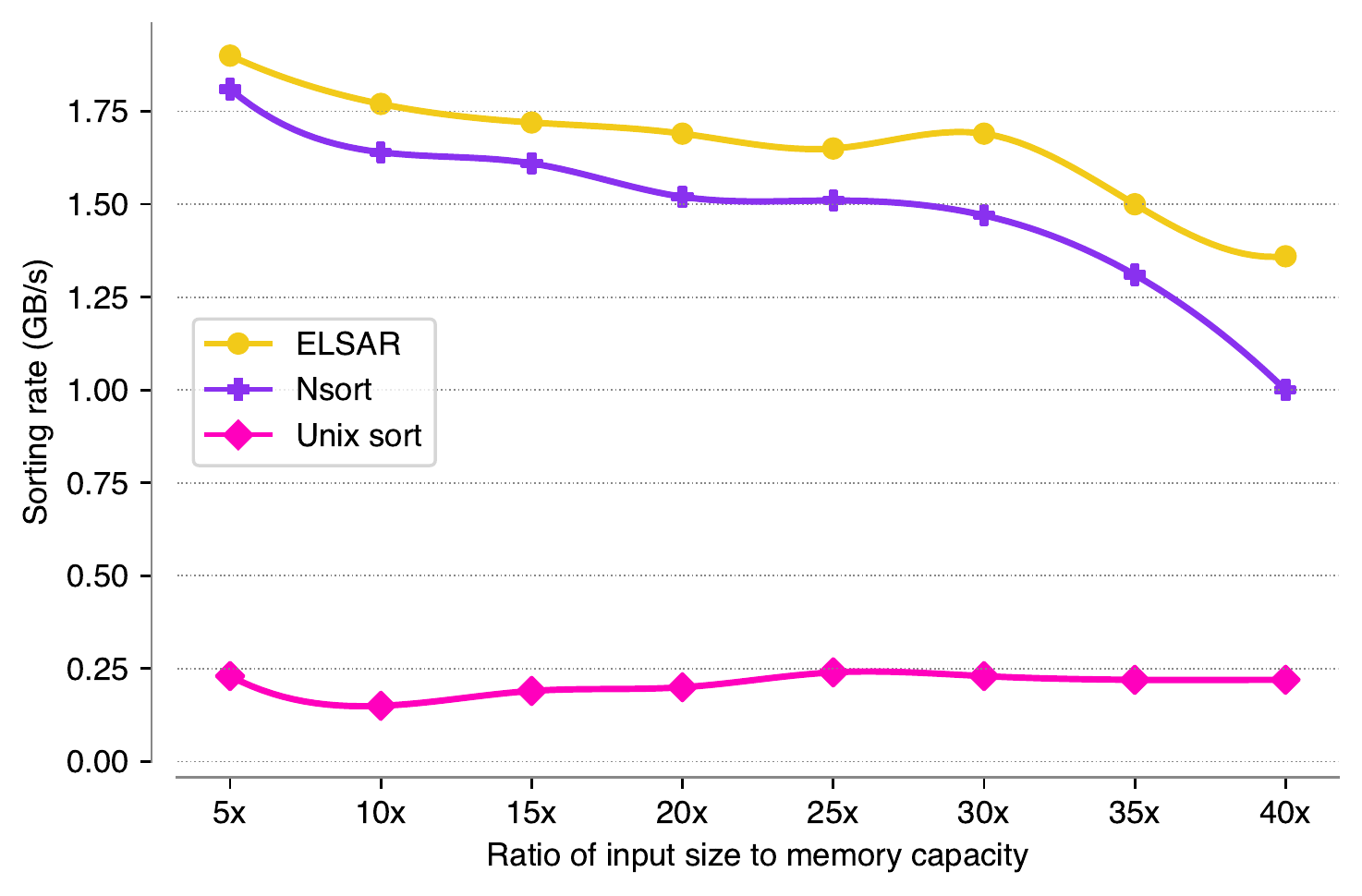}
    \caption{The scalability experiment showing the performance of \algo and other baselines as the input size gets progressively larger (higher is better). 
    The evaluation is performed on the Aurora machine, on the non-skewed datasets varying from 150~GB and up to 1.2 TB.
    The average disk bandwidth on the Aurora machine is 2.93~GB/s. }
    \label{fig:data_scalability} 
\end{figure}

Next, we look at how these algorithms scale with increasing input sizes. 
Therefore, we compare the sorting rates of \algo for inputs sizes starting from $5\times$ and up to $40\times$ the size of physical memory. 
Figure \ref{fig:data_scalability} shows the collected results.

We performed these measurements on the Aurora machine, using 10 hardware threads. 
The disks are M.2 NVMe\TM SSDs that can reach sequential read speeds up to 7.00~GB/s\cite{wdblack_specs}.  
Based on our measurement, as described in the section above, a parallel, complete input read and immediate write to these disks results in a throughput of 2.93~GB/s. 
This represents an almost $3\times$ higher throughput from the NVMe\TM disks used by the McGraw machines, hence the difference in observed speeds of the algorithms in Figures~\ref{fig:storage_types} and \ref{fig:data_scalability}.

We observed that the average sorting rate of \algo is 1.66~GB/s, Nsort's is 1.48~GB/s, and Unix sort's is 0.21~GB/s. 
At each increment of $5\times$ (i.e., 150~GB), \algo slows down by 5\% on average, whereas Nsort 8\%. 
Thus, when the input size is $40\times$ that of the memory capacity, Nsort's throughput has dropped by 45\% compared to the starting point, while \algo's only by 28\%. 
Finally, Unix sort's performance is relatively steady, only dropping, on average, by 1\% at each step and 4\% in total. 

\subsection{Energy efficiency}

\begin{figure}
    \includegraphics[width=\linewidth]{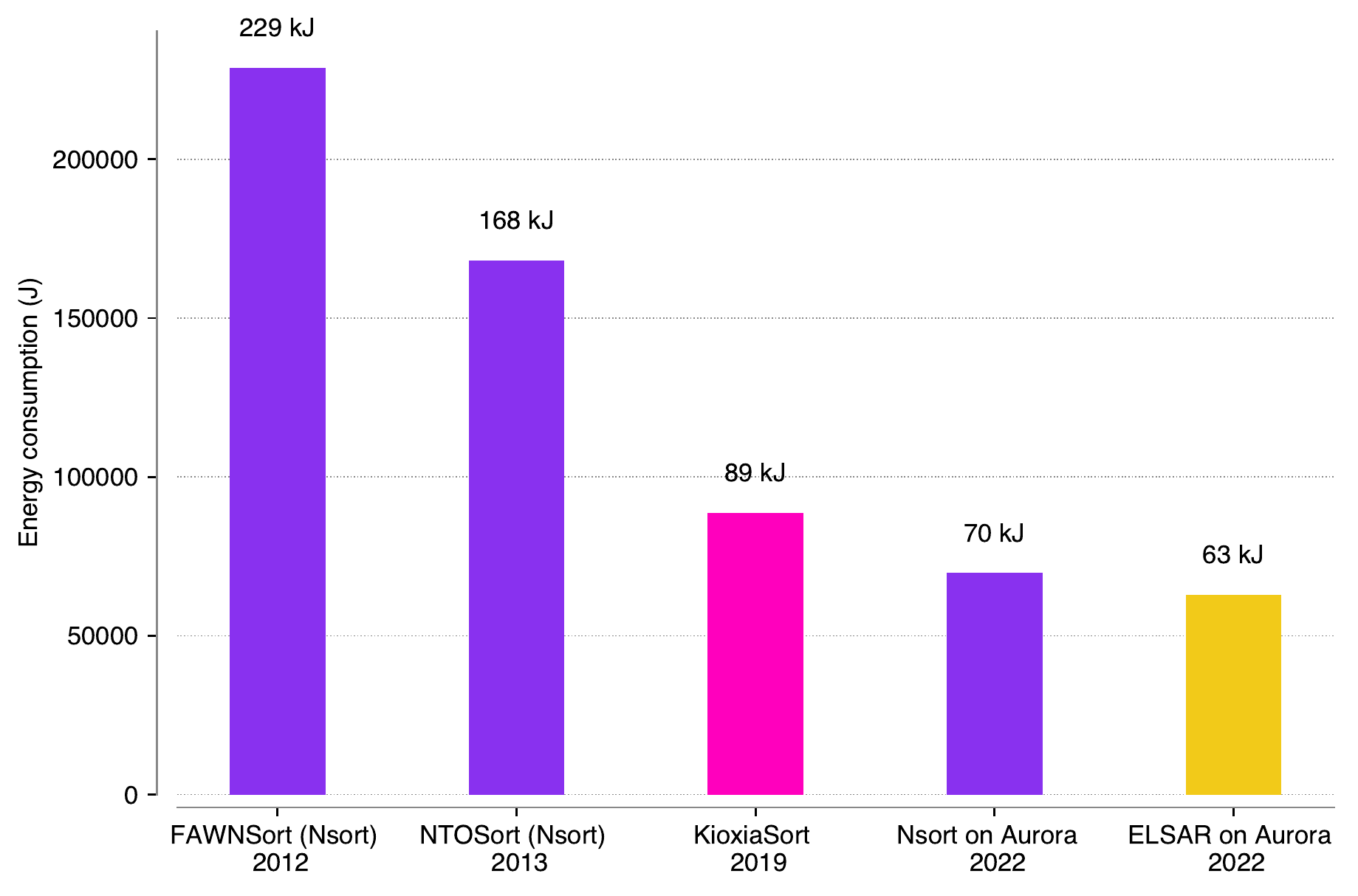}
    \caption{The energy efficiency of \algo and Nsort on the Aurora machine, and the past three winners of the SortBenchmark JouleSort category dating back up to 2012 (lower is better). 
    \algo's total energy consumption on the Aurora machine is 63~kJ, which is 41\% lower than the current benchmark winner - KioxiaSort\cite{kioxia_sort} at 89~kJ.
    KioxiaSort is an External Mergesort algorithm.
    The FAWNSort and NTOSort submissions both use Nsort but on different hardware. 
    We show the energy consumption of Nsort on our hardware (Aurora) as well for comparison. 
    }
    \label{fig:energy} 
\end{figure}

Another perspective for evaluating external sorting algorithms is their energy efficiency. 
Energy consumption is calculated as the product of the total execution time and the system's average power. 
For example, a sorting algorithm that runs on a system whose CPU has a massive number of cores to boost parallelism or connected to many fast disks to increase I/O bandwidth may perform extremely fast. 
However, its power consumption is also very high. 
For example, the PMem drive used in the McGraw machine has $1.4\times$ faster performance than the NVMe\TM SSD, but its average power rating is $3\times$ higher\cite{optane_specs,mcgraw_nvme_specs}.
In contrast, the algorithm might be slower on a desktop system with a limited number of CPU cores and disks, but its overall power is low. 
For example, typical desktop-grade processors, such as the Intel\R Core\TM i5 with ten cores, has a base TDP of 65~W, which is $1.92\times$ lower than that of the Intel\R Xeon\R Gold 6230 processor used in the McGraw machine, which has 20 cores.
Therefore, the energy consumption metric is an excellent choice to compare external sorting algorithms on different hardware because it incorporates both the time and system resources via the power dissipation measure.
In addition, we do not possess the source code or binaries of the sorting algorithms in Figure~\ref{fig:energy}, so we can only compare their reported energy readings. 

The SortBenchmark has a particular category - JouleSort\cite{joulesort}, dedicated to submissions for the most energy-efficient algorithm.   
The importance of energy efficiency is associated with lower business costs for providing power to the computer systems running these processes, lower costs for building adequate cooling infrastructure, and environmental and sustainability concerns. 
Each year, the benchmark accepts submissions that have been able to sort 1TB of 100-byte ASCII records with 10-byte keys while consuming less energy than the existing record. 
The current winner of the benchmark is a 2019 submission named KioxiaSort\cite{kioxia_sort}, which performs the task with 89~kJ of energy.
Since the metric is energy consumption and not only execution time, entries are allowed to use any hardware configuration, with the majority using desktop-grade machines\cite{kioxia_sort,nto_sort,fawn_sort,dem_sort}. 
Other SortBenchmark categories like GraySort and CloudSort contain submissions from very large cloud providers like Tencent\cite{tencent_sort} or Alibaba\cite{nad_sort}, who use high-speed networks of several hundred virtual machines to sort 100~TB of records. 
While we would have loved to participate in those benchmarks as well, it appears that cloud pricing and the ability to have access to huge amounts of servers are more critical in those categories than the actual sorting techniques used. 
Figure~\ref{fig:energy} summarizes the energy consumption of the last three winners of the SortBenchmark JouleSort category, as well as the energy consumption of Nsort and \algo on the Aurora machine, the machine we purposefully build to be energy efficient for sorting.  

In Figure~\ref{fig:energy}, the first bar belongs to FAWNSort (2012), which uses Nsort on a system with Intel\R Core\TM i7 with 8~GB of RAM and 16 SSDs\cite{fawn_sort}. 
The second bar belongs to NTOSort (2013), which also uses Nsort, and their system also consists of a newer-generation i7 processor and 16 faster SSDs\cite{nto_sort}. 
The third one is KioxiaSort (2019), which implements its version of an External Mergesort on a system with an Intel\R Core\TM i9 processor with 64~GB of RAM and 8 RAID-0 NVMe\TM SSDs. 
Finally, the last two bars show the energy consumption of Nsort and \algo on our Aurora machine. 

\algo on the Aurora machine consumes 62,912~J, which is 41\% lower than the current winner of the benchmark. 
This sets a new candidate for the record for the lowest energy consumption for the benchmark task by a significant margin.

For reference, we also show the energy consumption of Nsort on Aurora, which is the same software used by FAWNSort and NTOSort, but on our hardware. 
We included the evaluation of Nsort on our machine to emphasize that the performance improvements seen by \algo come as a result of the algorithmic design and not just purely from the hardware used. 
Nsort uses 69,772~J, which is 11\% larger than \algo's energy consumption. 

\subsection{A breakdown of the algorithmic portions}

\begin{figure}
    \includegraphics[width=\linewidth]{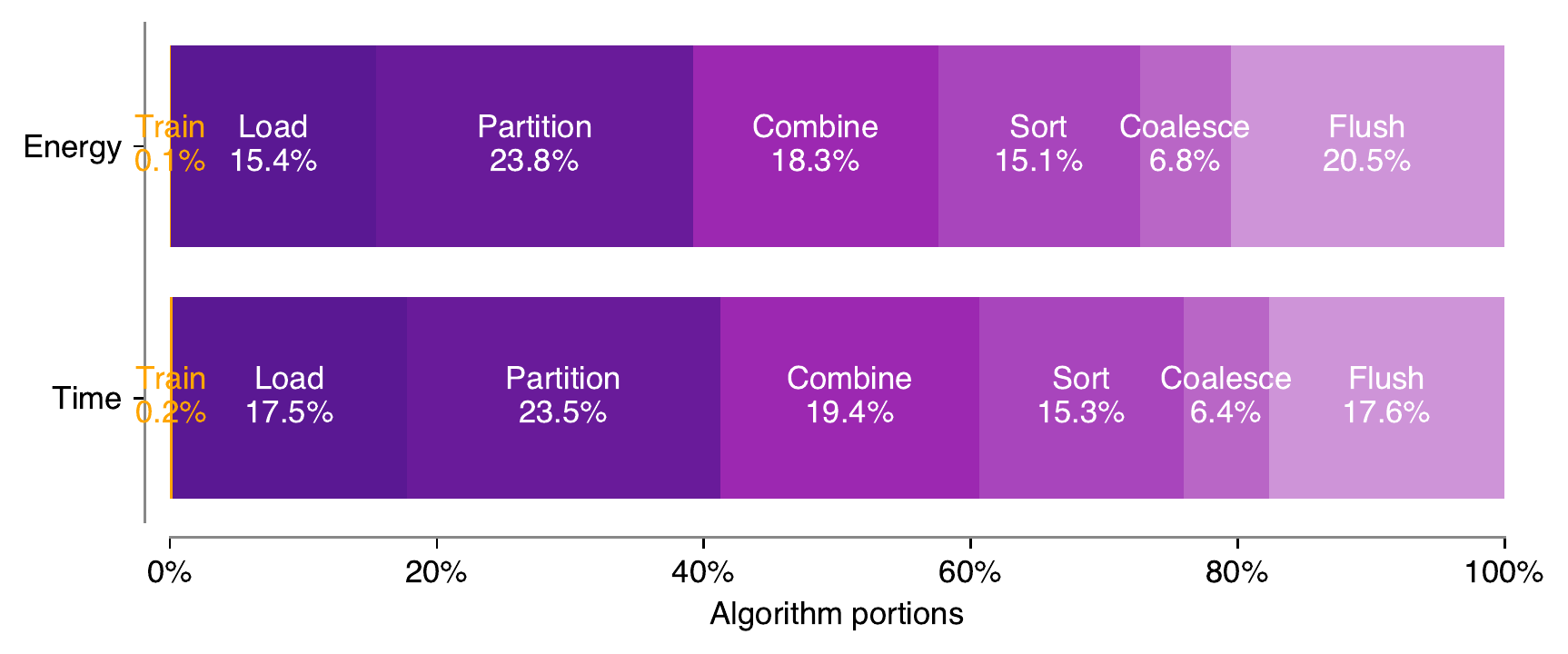}
    \caption{The breakdown of the algorithmic portions of \algo in terms of execution time and energy consumption. 
    These measurements were performed on the Aurora machine. 
    The energy consumption of each portion was calculated as the number of seconds consumed by it multiplied by the average power consumed by the algorithm up until then. 
    The training time and energy are not visible, representing less than 1\% (in orange for contrast). 
    }
    \label{fig:break-down} 
\end{figure}

We analyze \algo even further by breaking down how much time the algorithm spends on each algorithm phase. 
In addition, we also look at the breakdown of energy consumption for each portion. 
The results are shown in Figure~\ref{fig:break-down}. 

The training procedure used to partition records is the smallest portion of the algorithm, taking <1\% of the total execution time and energy consumption.
On the other hand, the biggest portion is the record partitioning (23.5\% of the total time and 23.8\% of the total energy). 
This is when the algorithm splits the input records onto different fragment files based on the CDF predictions. 
It is the most expensive operation of the algorithm as the records are not in any particular order, and they incur random writing patterns. 

In addition, record coalescing takes up roughly 7\% of the time and energy. 
We added this optimization to speed up the time it takes the algorithm to flush the records to the output file by making the writes in sequential batches. 
While 7\% is not insignificant, the total time taken by coalescing and flushing ($
\sim 24\%$) is still less than how much it would take to perform the flushing without coalescing.

\subsection{I/O performance}

\begin{figure}
    \includegraphics[width=\linewidth]{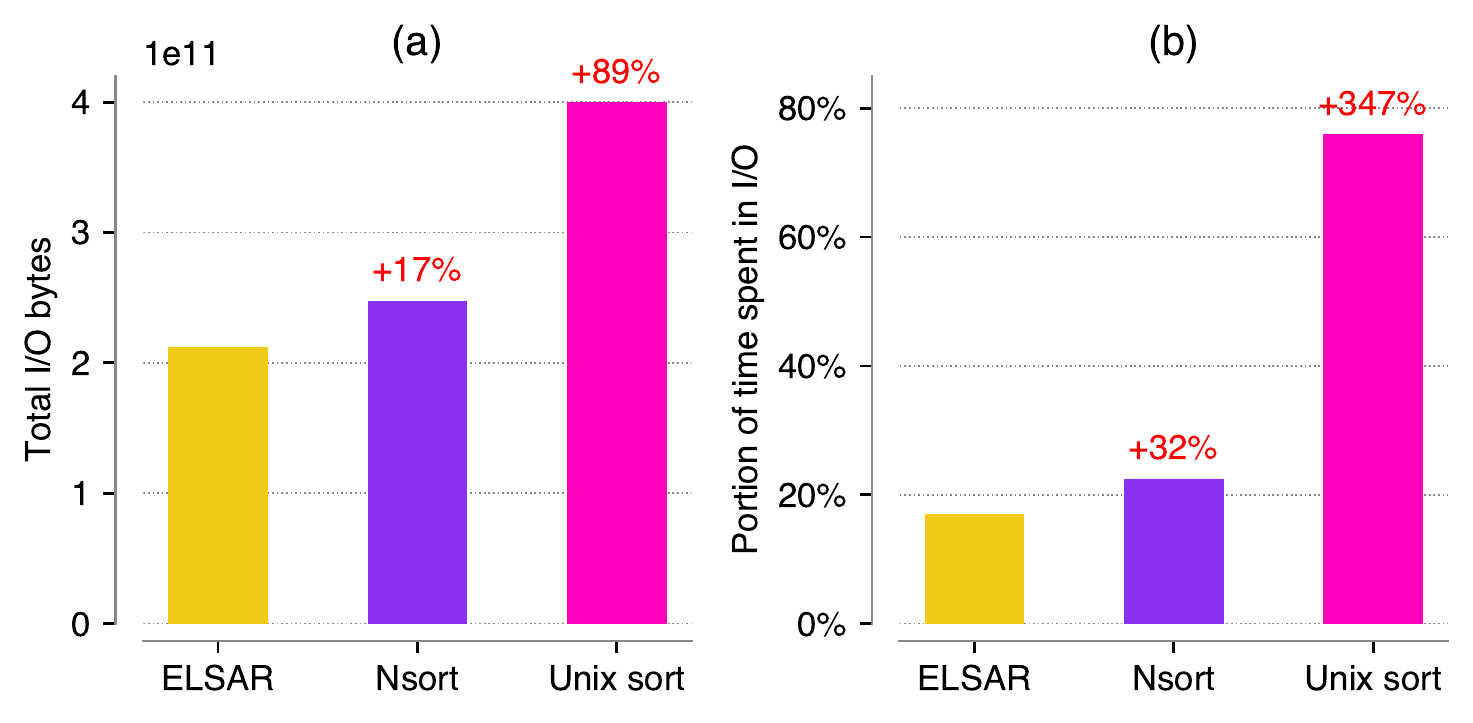}
    \caption{The I/O performance of \algo and other baseline algorithms. 
    In (a), we show the I/O size of each algorithm, defined as the total number of bytes written and read to/from disk (lower is better). 
    In (b), we show the portion of the time that these algorithms spend performing \texttt{read} and \texttt{write} system calls.}
    \label{fig:io} 
\end{figure}

Finally, we also dissect the I/O performance of \algo, Nsort, and Unix sort by looking at the I/O workload of each algorithm and the portion of execution time they spend doing I/O.
This analysis allows us to quantify how much these algorithms are disk I/O-bound, which is the major bottleneck of external sorting. 
In order to provide this metric, we use the \texttt{strace} Linux utility to log the size of the \texttt{read} and \texttt{write} system calls of each algorithm and their children processes (i.e., threads). 
We perform this evaluation on the Aurora machine with an input of size 100GB. 

In Figure~\ref{fig:io}(a), we show the I/O load of each algorithm, which we define as the total number of bytes written to and read from disk, as reported by the \texttt{strace} tool. 
Relative to \algo, Nsort's I/O load is 17\% larger, and Unix sort's is 89\% larger. 

On the other hand, in Figure~\ref{fig:io}(b), we show the portion of the execution time that these algorithms spend performing I/O operations on disk. 
\algo spends 17\% of its total execution time performing I/O, whereas Nsort spends 23\% of its time in I/O, which is 32\% longer than \algo. 
Finally, for Unix sort, the I/O time is 76\% of its total execution time, which is almost $3.5\times$ higher than \algo's. 

These results show that \algo spends less time performing I/O due to a smaller load, leading to better overall performance.

\section{Conclusion}
This paper introduced a novel external sorting algorithm, \algo, which leverages learned data distribution models. 
Unlike the external merge-sort paradigm, \algo partitions the input in a mutually exclusive, equi-depth, and monotonically increasing way, eliminating the need for a file merging routine.
Instead, it performs a simple file concatenation, which is significantly faster. 
We also showed multiple evaluations of \algo and existing external sorting algorithms on different machines, storage media, distributions, and input sizes. 
We consistently observed higher sorting rates for \algo than other sorting algorithms.
Furthermore, \algo superseded the SortBenchmark JouleSort record for the most energy-efficient sorting algorithm by 41\%.  

Nevertheless, this work only represents the initial results of this sorting algorithm. 
In the future, we intend to make \algo a high-performing distributed sorting algorithm that can work with datasets in the order of hundreds of terabytes and also supports Unicode characters.

\balance

\bibliographystyle{ACM-Reference-Format}
\bibliography{bib}

\noindent{\small Optimization Notice: Software and workloads used in performance tests may have been optimized for performance only on Intel microprocessors. Performance tests, such as SYSmark and MobileMark, are measured using specific computer systems, components, software, operations and functions. Any change to any of those factors may cause the results to vary. You should consult other information and performance tests to assist you in fully evaluating your contemplated purchases, including the performance of that product when combined with other products. For more information go to http://www.intel.com/performance. Intel, Xeon, and Intel Xeon Phi are trademarks of Intel Corporation in the U.S. and/or other countries.}

\end{document}